# Inset Edges Effect and Average Distance of Trees

M. H. Khalifeh*, A.-H. Esfahanian

Department of Computer Science and Engineering Michigan State University East Lansing, MI 48824, USA

**Abstract:** An added edge to a graph is called an *inset edge*. Predicting $k$ inset edges which minimize the average distance of a graph is known to be NP-Hard. When $k = 1$ the complexity of the problem is polynomial. In this paper we further find the single inset edge(s) of a tree with the closest change on the average distance to a given input. To do that we may require the effect of each inset edge for the set of inset edges. For this, we propose an algorithm with the time complexity between $O(m)$ and $O(m\sqrt{m})$ and average of less than $O(m.\log(m))$, where $m$ stands for the number of possible inset edges. Then it takes up to $O(\log(m))$ to find the target inset edges for a custom change on the average distance. Using theoretical tools, the algorithm strictly avoids recalculating the distances with no changes, after adding a new edge to a tree. Then reduces the time complexity of calculating remaining distances using some matrix tools which first introduced in [8] with one additional technique. This gives us a dynamic time complexity and absolutely depends on the input tree which is proportion to the Wiener index of the input tree.

# Introduction:

Average distance, degree distribution and clustering coefficient are the three most robust measures of network topology. The average distance of a graph is the average of distances between every pair of vertices with finite distances [2].

---

*Corresponding author's Email: khalife8@msu.edu

A proper change on the average distance of a network after missing or forming a link is interesting for the researchers [3-5, 11]. The efficiency of mass transfer in a metabolic network can be judged by studying its average path length [6]. If we add a new edge to a graph, we call the added edge an *inset edge*. Predicting $k$ inset edges which minimize the average distance of a graph is an NP-Hard problem [13]. This faces us a big challenge to find an inset edge with a given change on the average distance. If we solve the problem for the trees, we might use it through a sampling using some random spanning trees of a graph.

In this paper using some tools from [8] with one additional technique an algorithm is designed. In the algorithm we sort the effect (the difference between the average distance of a tree and the graph formed after adding an edge to the tree) of each inset edge for the set of inset edges of a tree. The algorithm in average is semi-linear. Note that a tree over $n$ vertices has $\frac{n^2-3n}{2} \cong O(n^2)$ inset edges which is the maximum possible inset edges among the set of $n$-vertex connected graphs.

## Definitions and Notations:

For a graph $G$ and $u, v \in V(G)$, $d_G(u, v)$ denotes the distance between $u$ and $v$ which is the length of a shortest path between $u$ and $v$ if there is a path between them and is equal to infinity otherwise. Using this for $A, B \subseteq V(G)$ let:

$$d_G(A, B) = \frac{1}{2} \sum_{(a,b) \in A \times B} d_G(a, b).$$

Thus, the sum of distances between all pair of vertices is,

$$D(G) = d_G(V(G), V(G)). \tag{1}$$

Therefore, the average distance of a graph is equal to $AD(G) = \frac{D(G)}{P}$ where $P$ stands for the number of pairs of vertices of $G$ with finite distances. When $G$ is connected, $P = \binom{|V(G)|}{2}$. $D(G)$ is known as Wiener index too [16].

In this work we analyze the possible effect of adding a new edge to the average distance of a given tree. For ease we may use $D(G)$ instead of $AD(G)$.

For a given graph $G \neq K_n$ (complete graph on $n$ vertices) and $Z \subseteq E(\overline{G})$, $_ZG = G + Z$. Therefore, for a tree $T$ and $xy \in E(T^c)$, $_{xy}T$ is a unicyclic graph. As mentioned, we call the edge $xy$ an *inset edge*. When we know the length of the cycle of $_{xy}T$ is $k$ we indicate it by $_{xy}T^k$.

**Definition 1:** For $_ZG$ we define:

$$D'(_ZG) = \frac{D(G) - D(_ZG)}{|Z|},$$

and

$$AD'(_ZG) = \frac{AD(G) - AD(_ZG)}{|Z|}.$$

Note that for $_{xy}T$, $|Z| = |\{xy\}| = 1$. Therefore $D'(T) = D(T) - D(_{xy}T)$. Moreover, for a connected graph $G$, $\binom{|V(G)|}{2} AD'(G) = D'(G)$.

Suppose $C$ is the cycle of $_{xy}T^k$. We define,

$$C_x = \{ v \in V(C) \mid d_T(x,v) < d_T(y,v) \},$$

$$C_y = \{ v \in V(C) \mid d_T(x,v) > d_T(y,v) \},$$

$$C_M = \{ v \in V(C) \mid d_T(x,v) = d_T(y,v) \}.$$

Note that $|C_x| = |C_y| = \left\lfloor \frac{k}{2} \right\rfloor$ and $|C_M| = 0$ when $k$ is even and $|C_M| = 1$ otherwise. Now we propose an indexing of $C_x$ to $x_i$'s, $1 \leq i \leq \left\lfloor \frac{k}{2} \right\rfloor$, such that $d_T(x, x_i) = \left\lfloor \frac{k}{2} \right\rfloor - i + 1$. Similarly, we index the elements of $C_y$ to $y_i$'s. Moreover, if $C_M \neq \phi$ then it has one element and we refer to that as $x_0$.

As we observe the indexing is unique and partitions the vertices of the cycle of $_{xy}T$. Using that, for $v \in V(C)$ suppose $T_v$ to be the maximal subtree of $_{xy}T$ such that,

$$V(T_v) \cap V(C) = \{v\}.$$

Informally $T_v$ is the tree attached to the vertex $v \in V(C)$. We denote the number of vertices of $T_v$, $|V(T_v)|$, by $w_v$.

Using the notations and definitions we will be using the following lemma from [8] which has a critical role for our analysis and reduction of the time complexity of actual calculations.

**Lemma 1 (see [8]):** For a tree $T$,

$$D'(_{xy}T^k) = \sum_{\substack{(u,v)\in C_x \times C_y \\ d_T(u,v) > \frac{k}{2}}} (2d_T(u,v) - k) \cdot w_u \cdot w_v$$

∎

As an interesting fact, we can see that:

$$max_{xy \in E(T^c)} D'(_{xy}T) = min_{xy \in E(T^c)} D(_{xy}T^k),$$

and also, by lemma 1:

$$D'(_{uv}T^3) = w_u \cdot w_v,$$

which is inspired by the derivative concept. We extend this idea using some matrix tools below. For more detail see [8]. For ease hereafter let $k' = \lfloor \frac{k}{2} \rfloor$.

Suppose we are given a $_{xy}T^k$. We associate the vectors $_{xy}^x W$ or $_{xy}^x W^k = [_{xy}^x w_i]$ to it which is a $k'$-vector and $_{xy}^x w_i = w_{x_i} = |T_{x_i}|$. Similarly, we define $_{xy}^y W$ or $_{xy}^y W^k = [_{xy}^y w_i]$ where $_{xy}^y w_i = w_{y_i} = |T_{y_i}|$. Finally, we associate a matrix $_{xy}W$ to $_{xy}T$ as follows,

$$_{xy}W = {_{xy}^x W} \times ({_{xy}^y W})^t.$$

Therefore, in a $_{xy}W = [w_{ij}]$, $w_{ij} = {_{xy}^x w_i} \cdot {_{xy}^y w_j} = w_{x_i} \cdot w_{y_j}$. Next, we introduce the matrix $F_k$. The matrix $F_k$ is a $k' \times k'$ matrix as follows,

$$F_k = \begin{cases} D_k + O_k & k \text{ is odd,} \\ D_k & \text{otherwise.} \end{cases}$$

where $D_k = [d_{ij}]$ and $O_k = [o_{ij}]$ are also $k' \times k'$ matrices as follows,

$$d_{ij} = \begin{cases} 2(k' - i - j + 1) & i + j \leq k', \\ 0 & \text{otherwise.} \end{cases}$$

and

$$o_{ij} = \begin{cases} 1 & i + j - 1 \leq k', \\ 0 & \text{otherwise.} \end{cases}$$

For more resolution,

$$F_k = \begin{bmatrix} 2k'-1 & 2k'-3 & 2k'-5 & & \cdots & & 1 \\ 2k'-3 & 2k'-5 & & & \cdots & 1 & 0 \\ 2k'-5 & & & \cdots & 1 & 0 & 0 \\ \vdots & & & & & & \vdots \\ & & \vdots & 1 & & & \\ \vdots & 1 & 0 & & & \cdots & 0 \\ 1 & 0 & 0 & & \cdots & 0 & 0 \end{bmatrix} \quad k \text{ is odd}$$

$$F_k = \begin{bmatrix} 2k'-2 & 2k'-4 & 2k'-6 & & \cdots & 2 & 0 \\ 2k'-4 & 2k'-6 & & & \cdots & 2 & 0 & 0 \\ 2k'-6 & \vdots & & \cdots & 2 & 0 & 0 & 0 \\ & \vdots & & & & & & 0 \\ & \vdots & 2 & & & & & \\ \vdots & 2 & 0 & & & & & \vdots \\ 2 & 0 & 0 & & & & \cdots & 0 \\ 0 & 0 & 0 & & & \cdots & 0 & 0 \end{bmatrix} \quad k \text{ is even}$$

$$O_k = \begin{bmatrix} 1 & 1 & 1 & & \cdots & & 1 \\ 1 & 1 & & & \cdots & 1 & 0 \\ 1 & & & \cdots & 1 & 0 & 0 \\ & & \vdots & & & & \vdots \\ & & & 1 & & & \\ \vdots & & 1 & 0 & & \cdots & 0 \\ 1 & & 0 & 0 & \cdots & 0 & 0 \end{bmatrix}$$

$$_{xy}W = \begin{bmatrix} w_{x_1} \\ w_{x_2} \\ \vdots \\ w_{x_{k'}} \end{bmatrix} \begin{bmatrix} w_{y_1} & w_{y_2} & \cdots & w_{y_{k'}} \end{bmatrix} = \begin{bmatrix} w_{x_1}.w_{y_1} & w_{x_1}.w_{y_2} & w_{x_1}.w_{y_3} & \cdots & w_{x_1}.w_{y_{k'}} \\ w_{x_2}.w_{y_1} & w_{x_2}.w_{y_2} & \vdots & \cdots & w_{x_2}.w_{y_{k'}} \\ w_{x_3}.w_{y_1} & \vdots & & \cdots & w_{x_3}.w_{y_{k'}} \\ \vdots & & \cdots & & \vdots \\ w_{x_{k'-2}}.w_{y_1} & \cdots & & \vdots & w_{x_{k'-2}}.w_{y_{k'}} \\ w_{x_{k'-1}}.w_{y_1} & \cdots & & \vdots & w_{x_{k'-1}}.w_{x_{k'-1}} & w_{x_{k'-1}}.w_{y_{k'}} \\ w_{x_{k'}}.w_{y_1} & \cdots & w_{x_{k'}}.w_{y_{k'-2}} & w_{x_{k'}}.w_{y_{k'-1}} & w_{x_{k'}}.w_{y_{k'}} \end{bmatrix}$$

We remind that if $A = [a_{ij}]$ is a matrix then the norm one of $A$ is the following:

$$\|A\| = \sum_{i,j} |a_{ij}|.$$

The Hadamard product of two matrices $A = [a_{ij}]$ and $B = [b_{ij}]$ with the same dimensions, $A \odot B = [c_{ij}]$, is an element-wise product with:

$$c_{ij} = a_{ij}.b_{ij}.$$

Also note that $e_i$ is the $i$th standard unique vector of proper dimension.

**Lemma 2 (see [8]):** For a tree $T$ we have:

$$D'(_{xy}T^k) = \|F_k \odot {}_{xy}W\|.$$

∎

**Definition 2:** Suppose $G$ is a graph and $u, v \in V(G)$ we define the relative neighborhood of $u$ and $v$ as follows:

$$N_G^v(u) = \{x \in V(G) \mid d(x,u) < d(x,v)\},$$

$$|N_G^v(u)| = n_G^v(u).$$

Note that if $T$ is a tree on $n$ vertices then for every $uv \in E(T)$, $n_T^v(u) + n_T^u(v) = n$.

**Remark 1:** The reason we propose the definition 2 is the fact that for $x_1, x_2 \in C_x$ and $y_1, y_2 \in C_y$ in $_{xy}T^k$,

$$n_T^{x_2}(x_1) = |T_{x_1}| = w_{x_1},$$

and

$$n_T^{y_2}(y_1) = |T_{y_1}| = w_{y_1},$$

which is required for our main algorithm.

The following algorithm produces the set $\{(n_T^v(u), n_T^u(v)) \mid uv \in E(T)\}$ which is required for our main algorithm.

### Algorithm 1:

**Input:** Adjacency list of a tree $T$ with $n$ vertices

Initiate with $R = \{v \in V(T) \mid \deg(v) = 1\}$ and $\{w(u) = 1, u \in R \text{ and } w(v) = 0, v \notin R\}$
While $(E(T) \neq \phi)$
    $S = R$
    $R = \phi$
    For $v \in S$
        If $(E(T) = \phi)$
            Break
        End
        $u = N(v)$
        $w(u) = w(u) + w(v)$
        $(n_T^v(u), n_T^u(v)) = (n - w(v), w(v))$
        $R = R \cup \{u\}$
        $E(T) = E(T) - uv$
    End
End

**Output:** $\{(n_T^v(u), n_T^u(v))\}_{uv \in E(T)}$

**Lemma 3:** For a tree $T$ on $n$ vertices we can obtain $\{(n_T^v(u), n_T^u(v))\}_{uv \in E(T)}$ in an $O(n)$ from its adjacency list. ▲

**Theorem 1:** Suppose $T$ is a tree and the vectors $_{xy}^x W$, $_{xy}^y W$ and the matrix $_{xy}W = [w_{ij}]$ are associated to $_{xy}T^k$ where $x$ and $y$ are not leaf. If $u \in N(x) - C_x$ and $v \in N(y) - C_y$ then

$$\begin{cases} D'(_{uv}T^{k+2}) - D'(_{xy}T^k) = -2 \sum_{i+j \le k'+1} w_{ij} + \sum_{i+j=k'+1} w_{ij} + 2. \sum_{r=2}^{k'} (_{uv}^{u}w_1 \cdot _{xy}^{y}w_r + _{xy}^{x}w_r \cdot _{uv}^{v}w_1) \\ \qquad -(_{uv}^{u}w_1 \cdot _{xy}^{y}w_{k'} + _{xy}^{x}w_{k'} \cdot _{uv}^{v}w_1), \quad k \text{ is odd}, \\ \\ D'(_{uv}T^{k+2}) - D'(_{xy}T^k) = -2 \sum_{i+j \le k'+1} w_{ij} + \sum_{i+j=k' \text{ or } k'+1} w_{ij} + 2 \sum_{r=2}^{k'} (_{uv}^{u}w_1 \cdot _{xy}^{y}w_r + _{xy}^{x}w_r \cdot _{uv}^{v}w_1) \\ \qquad -2(_{uv}^{u}w_1 \cdot _{xy}^{y}w_{k'} + _{xy}^{x}w_{k'} \cdot _{uv}^{v}w_1), \quad k \text{ is even}, \end{cases}$$

**Proof:** Suppose we have $_{xy}T^k$ with $d_T(x,y) > 2$. With the theorem condition, if we remove $xy$ from $_{xy}T^k$ and add $uv$ to $T$ where $u \in N(x) - C_x$ and $v \in N(y) - C_y$ we have:

$$_{uv}^{u}W = [n_T^x(u), _{xy}^{x}w_1, _{xy}^{x}w_2, \dots, _{xy}^{x}w_{k'}] - n_T^x(u).e_2 \qquad (1)$$

and

$$_{uv}^{v}W = [n_T^y(v), _{xy}^{y}w_1, _{xy}^{y}w_2, \dots, _{xy}^{y}w_{k'}] - n_T^y(v).e_2 \qquad (2)$$

Expanding $_{uv}W = {}_{uv}^{u}W \times {}_{uv}^{v}W$ using (1) and (2) through lemma 2, $D'(_{uv}T^{k+2}) - D'(_{xy}T^k) = \|F_{k+2} \odot {}_{uv}W\| - \|F_k \odot {}_{xy}W\|$ that is:

$$\begin{cases} D'(_{uv}T^{k+2}) - D'(_{xy}T^k) = -2 \sum_{i+j \le k'+1} w_{ij} + \sum_{i+j=k'+1} w_{ij} + 2. \sum_{r=2}^{k'} (_{uv}^{u}w_1 \cdot _{xy}^{y}w_r + _{xy}^{x}w_r \cdot _{uv}^{v}w_1) \\ \qquad -(_{uv}^{u}w_1 \cdot _{xy}^{y}w_{k'} + _{xy}^{x}w_{k'} \cdot _{uv}^{v}w_1), \quad k \text{ is odd}, \\ \\ D'(_{uv}T^{k+2}) - D'(_{xy}T^k) = -2 \sum_{i+j \le k'+1} w_{ij} + \sum_{i+j=k' \text{ or } k'+1} w_{ij} + 2. \sum_{r=2}^{k'} (_{uv}^{u}w_1 \cdot _{xy}^{y}w_r + _{xy}^{x}w_r \cdot _{uv}^{v}w_1) \\ \qquad -2(_{uv}^{u}w_1 \cdot _{xy}^{y}w_{k'} + _{xy}^{x}w_{k'} \cdot _{uv}^{v}w_1), \quad k \text{ is even}, \end{cases}$$

This completes the proof. ■

**Definition 3:** Each inset edge connects two vertices of a tree. The path connects the two vertices either has a middle vertex or a middle edge. We call the middle

vertex(edge) the **middle** of that inset edge. A vertex(edge) of a tree can be the middle of more than one inset edge.

**Theorem 2:** The set of inset edges of a tree forms a partition regarding the middles. More precisely each class of the partition is the inset edges of the tree with the same middles.

**Proof:** Since the length of each path is fixed so the middles are fixed. On the other hand, one inset edge in a tree forms a unique cycle in the tree. Therefore, every inset edge has a unique middle. These complete the proof. ∎

## A Matrix Norm Technique

In general, the time complexity of calculating the norm of a square $k \times k$ matrix is $O(k^2)$. For a given tree, $T$, by lemma 2 to calculate $D'(_{xy}T^k)$ we require to calculate $\|F_k \odot {}_{xy}W\|$ with a general time complexity of $O(k^2)$. Here we propose a technique, using the regularity of $F_k \odot {}_{xy}W$, to reduce the total time complexity of calculating $\{D'(_{xy}T^k)\}_{xy \in E(G)}$. But note that:

I- Using lemma 2 it is not possible to directly calculate $D'(_{xy}T^k)$ in less than $O(k^2)$.
II- We can achieve $D'(_{x_i y_i}T^{k+2})$ from $D'(_{x_{i+1}y_{i+1}}T^k)$ in $O(k)$, $1 \leq i \leq k'$.

From now if $A$ is a given matrix we will use $A[spesificatin\ of\ indices]$ which denotes a matrix correspond to $A$ where specified entries are equal to the matrix $A$ and are zero otherwise.

**Proposition 1:** Suppose $T$ is a tree, $xy \in V(T^c)$, $u \in N(x) - C_x$ and $v \in N(y) - C_y$. If we have ${}_{xy}W[i+j \leq k'+1]$, $D'(_{xy}T^k)$, $\|{}_{xy}W[i+j \leq k'+1]\|$ and also $w_u$ and $w_v$ regarding ${}_{uv}T^{k+2}$. Then we can obtain $D'(_{uv}T^{k+2})$, ${}_{uv}W[i+j \leq k'+2]$ and $\|{}_{uv}W[i+j \leq k'+2]\|$ in an $O(k)$.

**Proof:** We prove the proposition for the odd $k$'s. Proof of the even cases is similar. Using the proposition assumptions and the theorem 1:

$$D'(_{uv}T^{k+2}) = D'(_{xy}T^k) - 2 \sum_{i+j \leq k'+1} w_{ij} + \sum_{i+j=k'+1} w_{ij} + 2 \cdot \sum_{r=2}^{k'} ({}_{uv}^u w_1 \cdot {}_{xy}^y w_r + {}_{xy}^x w_r \cdot {}_{uv}^v w_1)$$

$$-({}_{uv}^{u}w_1 \cdot {}_{xy}^{y}w_{k'} + {}_{xy}^{x}w_{k'} \cdot {}_{uv}^{v}w_1)$$

$$= D'({}_{xy}T^k) - 2\|{}_{xy}W[i+j \leq k'+1]\| + \|{}_{xy}W[i+j = k'+1]\|$$
$$+ w_u \sum_{r=1}^{k'} {}_{xy}^{y}w_r + w_v \sum_{r=1}^{k'} {}_{xy}^{x}w_r \qquad (7)$$

According to the theorem we are given the $D'({}_{xy}T^k)$ and $\|{}_{xy}W[i+j \leq k'+1]\|$. Moreover, we clearly are able to calculate $2\|{}_{xy}W[i+j = k'+1]\|$ and $w_u \sum_{r=1}^{k'} {}_{xy}^{y}w_r$ and $w_v \sum_{r=1}^{k'} {}_{xy}^{x}w_r$ in $O(k)$ and so $D'({}_{uv}T^{k+2})$ from (7) in $O(k)$. To obtain ${}_{uv}W[i+j \leq k'+1]$ we can obtain the vectors $w_u[w_v, {}_{xy}^{y}w_1, \ldots, {}_{xy}^{y}w_{k'}]$ and $w_v[w_u, {}_{xy}^{x}w_1, \ldots, {}_{xy}^{x}w_{k'}]^T$ from ${}_{xy}W[i+j \leq k'+1]$ and add to ${}_{xy}W[i+j \leq k']$ properly. This takes $O(k)$ operations using the first row and column of ${}_{xy}W[i+j \leq k']$. Finally, using the theorem assumptions, the LHS of the following equality can be reached in $O(k)$.

$$\|{}_{uv}W[i+j \leq k'+2]\| = \|{}_{xy}W[i+j \leq k'+1]\| - \|{}_{uv}W[i+j = k'+1]\|$$
$$+ 2\left\|w_u\left[{}_{xy}^{y}w_1, \ldots, {}_{xy}^{y}w_{k'}\right]\right\| + 2\left\|w_v\left[{}_{xy}^{x}w_1, \ldots, {}_{xy}^{x}w_{k'}\right]\right\|.$$

This completes the proof. ∎

# Algorithm discussion

In this section using our result, we first propose the algorithm 2 which is recursive. Then we discuss that. The algorithm calculates the $D'$ index of the set of inset edges of a given tree.

### Algorithm 2:

**Input:** a tree T with $\{(n_T^v(u), n_T^u(v))\}_{uv \in E(T)}$

For $v \in V(T)$:
    For $a \in N(v)$
      For $b \in N(v)$
        if $(a \neq b)$

$$W = [n_T^v(a) . n_T^v(b)]$$
$$A = N(a) - a$$
$$B = N(b) - b$$
$$V_x = [n_T^v(a)]$$
$$V_y = [n_T^v(b)]$$
$$WU = \|W\|$$
$$D' = \|W\|$$
**Out** $D'$
$$FV(W, WU, V_x, V_y, D, A, B, x, y, k = 1)$$
    End
  End
End

$FV(W, WU, V_x, V_y, D, A, B, x, y, k = 1)$
if $(A \neq \emptyset \wedge B \neq \emptyset)$
 For $a \in A$
   For $b \in B$
     $A = N(a) - x$
     $B = N(b) - y$
     $V_x = [n_T^x(a)] \wedge V_x$
     $W \to$ Append $V_x$ as top row of $W$
     $W \to$ Append $V_y$ as left column of $W$
     $W[2:] = W[2:] - n_T^y(b) \, V_x$
     $W[2,2] = n_T^x(a) . n_T^y(b)$
     $V_y = [n_T^y(a)] \wedge V_y$
     $W[:2] = W[:2] - n_T^x(a) \, V_y$
     $D' = D' - 2WU + \|W[i+j = k'+1]\| + 2n_T^y(b)\|V_x\| + 2n_T^x(a)\|V_y\|$
        $-n_T^y(b) \, V_x[k'] - n_T^x(a) \, V_y[k'] - 4n_T^x(a) . n_T^y(b)$
     $WU = WU - \|W[i+j = k'+1]\| - n_T^y(b) \, V_x[-1] - n_T^x(a) \, V_y[-1]$
     **Out** $D'$
    **return** $FV(W, WU, V_x, V_y, D, A, B, x, y, k = k+1)$
   End
  End
End

For $uv \in E(T)$:
    For $a \in N(u) - v$
       For $b \in N(v) - u$
$$W = \begin{bmatrix} n_T^v(a) \\ n_T^u(v) - n_T^v(a) \end{bmatrix} . [n_T^u(b) \quad n_T^v(u) - n_T^u(b)]$$
           $A = N(a) - a$
           $B = N(b) - b$

$$V_x = [n_T^u(a)]$$
$$V_y = [n_T^v(b)]$$
$$WU = \|W[i+j \neq 4]\|$$
$$D' = 2\|W\|$$
**Out** $D'$
$$FE(W, WU, V_x, V_y, D, A, B, x, y, k=1)$$
    End
  End
End

$FE(W, WU, V_x, V_y, D, A, B, x, y, k=1)$

if $(A \neq \emptyset \wedge B \neq \emptyset)$
 For $a \in A$
  For $b \in B$
    $A = N(a) - x$
    $B = N(b) - y$
    $V_x = [n_T^x(a)] \wedge V_x$
    $W \to$ Append $V_x$ as top row of $W$
    $W \to$ Append $V_y$ as left column of $W$
    $W[2:] = W[2:] - n_T^y(b) V_x$
    $W[2,2] = n_T^x(a) \cdot n_T^y(b)$
    $V_y = [n_T^y(a)] \wedge V_y$
    $W[:2] = W[:2] - n_T^x(a) V_y$
    $D' = D' - 2WU + 2\|W[i+j = k'+1]\| + 2\|W[i+j = k'+2]\|$
        $+ 2n_T^y(b)\|V_x\| + 2n_T^x(a)\|V_y\| - n_T^y(b) V_x[k] - n_T^x(a) V_y[k] - 4n_T^x(a) \cdot n_T^y(b)$
    $WU = WU - \|W[i+j = k'+1]\| - n_T^y(b) V_x[-1] - n_T^x(a) V_y[-1]$
    **Out** $D'$
    **return** $FE(W, WU, V_x, V_y, D, A, B, x, y, k=k+1)$
    End
  End
End
**Output**: $\{D'(_{xy}T^k)\}_{xy \in E(T^c)}$

## Algorithm discussion

Every vertex(edge) is a middle for a class of inset edges. See the Theorem 2. As the algorithm 2 presents we start from every vertex and edge of a tree. Then the related average distances' change of a class of inset edges will be calculated. The calculations are based on the Proposition 1.

**Lemma 3:** Suppose $v$ or $e$ is the middle of the inset edge, $xy$, of a tree $T$, with $d(x, v \text{ or } e) = d(y, v \text{ or } e) = k$. Then the algorithm achieves $D'(_{xy}T^k)$ in $O(k)$.

**Proof:** This is enough to look at the algorithm 2 and Proposition 1. ∎

**Lemma 4 (see [8,15]):** If $T$ is a tree on $n$ vertices then:

$$(n-1)^2 \leq D(T) \leq \binom{n+1}{3},$$

and

$$1 \leq D'(_{xy}T) \leq \frac{n^3}{16} - \frac{n^2}{32} - \frac{9n}{8} + 2.$$

∎

As wiener index of *chemical graphs*, $D$, correlates with certain physical and chemical properties of molecules, there are several efficient algorithm for calculating the wiener index and generally distance-based invariants of graphs [1,7,10,12]. Using [9] for a tree $T$,

$$D(T) = \sum_{uv \in E(T)} n_T^v(u) \cdot n_T^u(v). \tag{8}$$

By the algorithm 1 which is linear for a tree on $n$ vertices, one can calculate $D(T)$ in $O(n)$ regarding (8).

Note that we cannot calculate $D'(_{xy}T^k)$ in $O(k)$ directly. The algorithm 2 achieves $D'(_{xy}T^k)$ using the $D'(_{x_2y_2}T^{k-2})$ through a recursive process. The next theorem gives an interesting relation between the complexity of computing $D'(_{xy}T)$'s and $D(T)$. Indeed, the algorithm 2s' complexity depends on the input tree.

**Theorem 3:** Suppose $T$ is a tree. The time complexity of calculating $\{D'(_{xy}T)\}_{xy \in E(T^c)}$ from the algorithm 2 is $O(D(T))$.

**Proof :** According to the lemma 3 if $xy$ is an inset edge and $d(x,y) = d$ then the algorithm 2 requires at most $c_1.d + c_2$ operations to calculate $D'(_{xy}T^k)$ for some constants $c_1$ and $c_2$. Therefore, we require at most

$$c_1 \sum_{xy \in E(T^c)} d_T(x,y) + \left(\binom{n}{2} - (n-1)\right) c_2 < c_1 D(T) + \left(\binom{n}{2} - (n-1)\right) c_2$$

operations to achieve $\{D'(_{xy}T^k)\}_{xy \in E(T^c)}$. Moreover, by lemma 4:

$$O\left(\binom{n}{2} - (n-1)\right) \leq O(D(T)).$$

This completes the proof. ∎

**Remark 2:** By the theorem 3 the complexity of computing $\{D'(_{xy}T^k)\}_{xy \in E(T^c)}$ is $O(D(T))$. Clearly $D(T) < \binom{n}{2} diam(T)$. If the average diameter of a tree on $n$ vertices is $\log(n)$ [14] then the average complexity of computing $\{D'(_{xy}T^k)\}_{xy \in E(T^c)}$ will be at most $O(n^2 \log(n))$. Since there are $O(n^2 = m)$ inset edges, we can sort $\{D'(_{xy}T^k)\}_{xy \in E(T^c)}$ in $O(m.\log(n))$. Therefore the algorithm 2, in average, sorts $\{D'(_{xy}T^k)\}_{xy \in E(T^c)}$ in no more than $O(m.\log m)$. And by the lemma 4 the worst case of algorithm will be $O(m\sqrt{m})$.

**Remark 3:** We saw that $D'(_{uv}T^3) = w_u.w_v$. It seems the complexity of computing each of $D'(_{xy}T^k)$'s for $xy \in E(T^c)$ in average can be constant which is already $O(d(x,y))$ by the algorithm 2. We probably require some more matrix techniques. Moreover, we guess that if we are only looking for the maximum $D'$ we are not required to consider all edges and vertices as middles where we run the algorithm 2. More precisely:

**Conjecture 1:** It is possible to calculate $\{D'(_{xy}T^k)\}_{xy \in E(T^c)}$ in $O(|E(T^c)|)$.

**Conjecture 2:** For a given tree, $T$, the middle(s) of the inset edge(s) with the maximum $D'$ belongs to $E(P') \cup V(P')$ with $P'$ is the longest path(s) between the $N(c)$ and $N(m_1) \cup N(m_2)$ where $c$ and $m_1 m_2$ are the center and median of $T$ respectively.

# References:

[1] Roberto Aringhieri, Pierre Hansen, Federico Malucelli, A Linear Algorithm for the Hyper-Wiener Index of Chemical Trees, J. Chem. Inf. Comput. Sci., 41 (2001), 958-963.

[2] F. Chung, L. Lu, The Average Distances in Random Graphs with Given Expected Degrees, Proceedings of the National Academy of Sciences, 99 (2002), 15879-15882.

[3] A. Clauset, C. Moore, M. E. J. Newman, Hierarchical Structure and the Prediction of Missing Links in Networks, Nature, 453 (2008), 98-101.

[4] J. Copic, M. O. Jackson, A. Kirman, Identifying Community Structures from Network Data via Maximum Likelihood Methods, The B.E. Journal of Theoretical Economics, 9 (2005), 09-27.

[5] S. Currarini, M. O. Jackson, P. Pin, An Economic Model of Friendship: Homophily, Minorities and Segregation, Econometrica, 77 (2009), 1003-1045.

[6] A. Drger, M. Kronfeld, M. J. Ziller, J, Supper, H. Planatscher, J.B. Magnus, M. Oldiges, O. Kohlbacher, A. Zell, Modeling metabolic net- works in C. glutamicum: a comparison of rate laws in combination with various parameter optimization strategies". BMC Systems Biology. doi:10.1186/1752-0509-3-5.

[7] Aleksander Kelenc, Sandi Klavžar, Niko Tratnik, The Edge–Wiener Index of Benzenoid Systems in Linear Time, MATCH Commun. Math. Comput. Chem., 74 (2015), 521-532.

[8] M. H. Khalifeh, A.-H. Esfahanian, Some Preliminary Result About the Inset Edge and Average Distance of Trees, submitted.

[9] M. H. Khalifeh, H. Yousefi Azari, A. R. Ashrafi, S. G. Wagner, Some new results on distance–based graph invariants, Eur. J. Comb. 30 (2009), 1149–1163.

[10] Sandi Klavžar, Petra Žigert, Ivan Gutman, An algorithm for the calculation of the hyper-Wiener index of benzenoid hydrocarbons, Computers & Chemistry, 24 (2000), 229-233.

[11] Y. Matsuo, Y. Ohsawa, M. Ishizuka, Average-Clicks: A New Measure of Distance on the World Wide Web, Journal of Intelligent Information Systems, 20 (2003), 20- 51.

[12] Bojan MOHAR, Tomaž Pisanski, How To Compute The Winer Index of a Graph, J. of Mathematical Chem., 267 (1988), 277 267.

[13] A. Meyerson, B. Tagiku, Minimizing Average Shortest Path Distances via Shortcut Edge Addition, Approximation, randomization, and com- binatorial optimization, Algorithms and techniques. 12th international workshop, APPROX 2009, and 13th international workshop, RANDOM 2009, Berkeley, CA, USA, August 2123, 2009. Proceedings (pp.272-285)

[14] Z. Shen, The average diameter of general tree structures, Comput. Math. Appl., 36 (1998), 111-130.

[15] H. B. Valikara, V. S. Shigehali, H. S. Ramane, Bounds on the Wiener number of a graph, 50 (2004), 117-132.

[16] H. Wiener, Structural determination of the paraffin boiling points, J. Am. Chem. Soc., 69 (1947), 17-20.